\documentclass[twocolumn,showpacs,preprintnumbers,amsmath,amssymb,epsfig]{revtex4}
\usepackage{epsfig,amssymb}
\usepackage{color}
\begin{document}

%\draft

\title{Phase Response Curves of Coupled Oscillators} 

\author{Tae-Wook Ko}
\email{taewook.ko@gmail.com}
\author{G. Bard Ermentrout}
\email{bard@math.pitt.edu}
\affiliation{Department of Mathematics, University of Pittsburgh, Pittsburgh, Pennsylvania 15260, USA}
\date{\today}
\begin{abstract}
Many real oscillators are coupled to other oscillators and the coupling can affect the response of the oscillators to stimuli.
We investigate phase response curves (PRCs) of coupled oscillators. The PRCs for two weakly coupled phase-locked oscillators are analytically obtained in terms of the PRC for uncoupled oscillators and the coupling function of the system. Through simulation and analytic methods, the PRCs for globally coupled oscillators are also discussed.  
% coupled oscillators, complex systems, neuroscience
%\\
\end{abstract}

\maketitle
%-----------------------------------------
Many systems in physics, chemistry and biology are modeled as
interacting nonlinear oscillators
\cite{pikovsky_book,kura,winfree,rinzel_bookchapt,ermentrout2001,tass_book}.
One of the easiest ways to characterize an oscillator is its
phase response curve (PRC)\cite{winfree,rinzel_bookchapt,ermentrout2001,tass_book,prc_papers}. 
The PRC  is defined as the steady phase shift of an oscillation relative to the unperturbed oscillation as a function of the timing of perturbation to the oscillator.
It provides a
useful information for understanding the oscillator's behavior when the oscillator is subjected to external stimuli or signals from other oscillators.  

In most of previous studies, the PRC is obtained when the oscillator is isolated from other oscillators \cite{winfree,rinzel_bookchapt,ermentrout2001,prc_papers}. However, many oscillators in real systems are coupled to others when they are under the influence of external stimuli, 
and the coupling can affect the response of the oscillators.
To better understand the dynamics of oscillators such as the response of neuronal population to signals from other brain region \cite{ermentrout2001} or to controlling stimulations \cite{tass_book}, it is necessary to study how the coupling changes the PRCs. 
This study can also give insights into the phase response of a giant oscillator (for example, circadian rhythm generators \cite{winfree}) composed of many individual oscillators \cite{kawamura2008}.
In this letter, we study the PRC of coupled oscillators
using the average phase of the system and the relative phases between the
oscillators comprising the system. 
The PRC is shown to depend on the PRC of the isolated oscillator, the nature of the coupling, and the relative phases between the oscillators.    
For some cases, the PRCs are analytically obtained.
Our approach differs from that of Ref. \cite{kawamura2008} in that we analytically approximate the PRC while they require the numerical evaluation of the adjoint of a certain linear operator.

If coupling between a network of oscillators is sufficiently ``weak'',
the possibly high-dimensional system can be reduced to a network
of coupled phase models \cite{kura,rinzel_bookchapt,ermentrout2001}.
 In the following we exploit
this fact and restrict our analysis to coupled phase models.
Consider, first,  two weakly coupled phase-locked oscillators
subjected to a common perturbation characterized by their individual
PRC: 
%---------------------------------
\begin{eqnarray}
\dot \theta_1 &=& \omega_1 + K H(\theta_2-\theta_1)+Z(\theta_1)\cdot A\delta(t-t_1),   
\label{eq_two_model01} \\
\dot \theta_2 &=& \omega_2 + K H(\theta_1-\theta_2)+Z(\theta_2) \cdot A\delta(t-t_1), 
\label{eq_two_model02}
\end{eqnarray}
 where $\theta_i (t)$ is the phase of oscillator $i$ at time $t$,
 $\omega_i$ is the natural frequency of the oscillator $i$ and $K
 (\geq 0)$ is the coupling strength. $H(\theta)$ is the coupling
 function obtained by the phase reduction
 \cite{kura,rinzel_bookchapt,ermentrout2001}. 
$A \delta(t-t_1)$ denotes a
 Dirac delta impulse with amplitude $A$ at time $t_1$ which is
 sufficiently large so that the perturbing impulse is applied after the
 system reaches a steady state.  $Z(\theta)$ is the PRC for uncoupled oscillator obtained using an impulse with unit amplitude. Without coupling ($K=0$), the impulse causes steady phase shift $AZ(\theta_i(t_1))$ for oscillator $i$.

In the presence of coupling ($K \neq 0$), 
if the oscillators are locked with nonzero phase difference, or the input
amplitudes are different, then the input impulse generally causes nonidentical phase changes to the oscillators. Thus, the system transiently deviates from the locked state and then returns to the state. The coupling can affect the phase shift which the oscillation of the recovered state can have relative to the unperturbed oscillation. 
 We wish to determine the PRC of the coupled oscillators, in other words, how the phase shift depends on the phase at $t_1$ of the perturbation. 

To analyze the dynamics, we convert Eqs. (\ref{eq_two_model01}) and
(\ref{eq_two_model02}) into those for the average phase $\Phi \equiv
\frac{\theta_1 + \theta_2}{2}$ and the relative phase $\phi \equiv
\theta_1 - \theta_2$. 
\begin{eqnarray}
\dot \Phi &=& \bar \omega + KH_{e}(\phi)+Z_{av}(\theta_1, \theta_2) \cdot A\delta(t-t_1),
\label{eq_phicm}\\
\dot \phi &=& \Delta \omega - 2 KH_{o}(\phi)+Z_d(\theta_1, \theta_2) \cdot A\delta(t-t_1), 
\label{eq_phid}
\end{eqnarray}
where
$\bar \omega = \frac{\omega_1 + \omega_2}{2}$, $\Delta \omega =
\omega_1 - \omega_2$, $Z_{av}(\theta_1, \theta_2) =
\frac{Z(\theta_1)+Z(\theta_2)}{2}$, $Z_d(\theta_1, \theta_2) =
Z(\theta_1)-Z(\theta_2)$, $H_{e}(\phi) = \frac{H(\phi)+H(-\phi)}{2}$,
and $H_{o}(\phi) = \frac{H(\phi)-H(-\phi)}{2}$. 

 For simplicity, let us assume that the system has one stable locked state with $\phi=\phi_0$ satisfying  $0=\Delta \omega - 2 KH_{o}(\phi_0)$ and ${H_{o}}'(\phi_0) > 0$. The phase of each oscillator can be written as 
$\theta_1 =  \Phi + \phi/2$ and $\theta_2 =  \Phi - \phi/2$. 
Let us denote the phase shift in a phase, for example $\theta_1$,
relative to the unperturbed oscillation by $\Delta \theta_1$. 
We can see that the phase shift for the
oscillator $1$ is given by    
\begin{eqnarray}
\Delta \theta_1 =  \Delta \Phi + \Delta \phi/2. 
\label{eq_Deltatheta_1}
\end{eqnarray}

When $H(\theta)$ is an odd function, the average phase $\Phi$ evolves
with a constant frequency $\bar \omega$ before and after the impulse
(Eq. (\ref{eq_phicm})).  
Thus, $\Delta \Phi = A Z_{av}(\theta_1(t_1), \theta_2(t_1))$.
 When the relative phase remains in the basin of attraction of the
 original relative phase right after the impulse, $\phi$ approaches the
 original relative phase. Otherwise, the relative phase moves to another
 stable value (called walkthrough). 
Thus, $\Delta \phi = \phi_f - \phi_0$, where $\phi_f$
 is the stable value of $\phi$ reached after the impulse. Note that even
 $\phi=\phi_0$ and $\phi=\phi_0 + 2\pi$ give different results.   
Therefore, the PRC of the oscillator $1$ in the coupled cases is given by
\begin{eqnarray}
{Z_{c1}}(\theta_1) =  A Z_{av}(\theta_1, \theta_1-\phi_0) + (\phi_f - \phi_0)/2.
\end{eqnarray}

%*************************************************
%Fig. 1
%\psdraft
\begin{figure}
\centering
\epsfig{figure=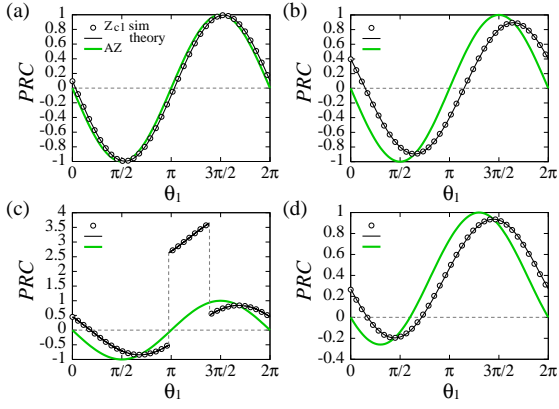, width= 7.5cm}
\caption{(Color online) PRCs with odd coupling functions. 
$\omega_1 = \pi/2 + \Delta \omega$ and $\omega_2 = \pi/2$. $\Delta \omega=0.4$ and $A=1$.
(a) $K=1$: $\phi_0 \approx 0.201$. 
(b) $K=0.25$: $\phi_0 \approx 0.927$.
(c) $K=0.22$: $\phi_0 \approx 1.141$.
For (a), (b), and (c),  $H(\theta)= \sin\theta$ and $Z(\theta)=-\sin\theta$.
(d) $H(\theta)= \sin\theta-0.4\sin(2\theta)$, $Z(\theta)=-\left[\sin(\theta+0.2\pi)-\sin(0.2\pi)\right]/\left[1+\sin(0.2\pi)\right]$, and $K=0.5$: $\phi_0 \approx 0.908$.
}
\label{fig_prc_oddH}
\end{figure}
%*************************************************
We simulate Eqs. (\ref{eq_two_model01}) and  (\ref{eq_two_model02}) using Euler method with time step $\Delta t = 0.01$.  
We measure the steady phase shift due to the impulse relative to the
unperturbed activity.  
The PRC is given by this phase shift as a function of the phase at
which the impulse is applied.

Figure \ref{fig_prc_oddH} shows $Z_{c1}(\theta_1)$ with odd coupling
functions. The prediction from the theory (black solid curves) matches
very well with the simulation results (symbols).  
With larger values of $\Delta \omega$ and/or smaller values of $K$,
the oscillators are locked with larger $\phi_0$. In
Figs. \ref{fig_prc_oddH}(a), (b), and (c) with $H(\theta)=\sin\theta$
and $Z(\theta)=-\sin\theta$, we show the PRC for different values of
the coupling strength $K$. When $\phi_0$ is very small, the PRC of
coupled oscillators is very close to that of uncoupled oscillators as
expected (Fig. \ref{fig_prc_oddH}(a)). In this case, $\phi$ goes to
the original value $\phi_0$ after the impulse. In
Fig. \ref{fig_prc_oddH}(b), with the larger $\phi_0$, the PRC of the
coupled oscillator becomes significantly different from that of
uncoupled oscillators.  
When the impulse can kick the system out of the basin of the stable
locked state with $\phi_0$, the system goes through phase walk
through. If
the system has a stable fixed point with $\phi_0$ and an unstable
fixed point $\phi_u$ in $[0, 2\pi)$ as in the case with
  $H(\theta)=\sin\theta$ for $\Delta \omega < 2K$, $\phi_u$ has the
  role of basin boundary and $\phi$ goes to $\phi_f=\phi_0+2\pi$ when
  $\phi_0 + AZ(\theta_1(t_1))-AZ(\theta_2(t_1)) > \phi_u$. This type
  of changes in $\phi$ causes the discontinuity shown in the PRC of
  Fig. \ref{fig_prc_oddH}(c).  
We show similar results for a coupling function with higher order Fourier
terms and an asymmetric PRC (Fig. \ref{fig_prc_oddH}(d)). 

%*************************************************
%Fig. 2
%\psdraft
\begin{figure}
\centering
\epsfig{figure=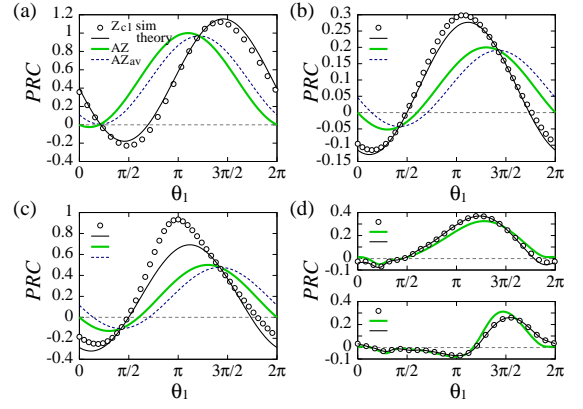, width= 7.5cm}
\caption{(Color online) PRCs with non-odd coupling functions.
For (a)-(c), $\omega_1 = \pi/2 + \Delta \omega$, $\omega_2 = \pi/2$ and $\Delta\omega=0.4$.
(a) $H(\theta)= \sin(\theta+0.4\pi)$, $Z(\theta)=-\left[\sin(\theta+0.4\pi)-\sin(0.4\pi)\right]/\left[1+\sin(0.4\pi)\right]$, and $K=1$: $\phi_0 \approx 0.704$, $A=1$.
(b) $H(\theta)= \sin(\theta-0.4\pi)+0.3\sin(2\theta-0.1\pi)$, $Z(\theta)=-\left[\sin(\theta+0.2\pi)-\sin(0.2\pi)\right]/\left[1+\sin(0.2\pi)\right]$, and $K=0.4$: $\phi_0 \approx 0.769$, $A=0.2$.
(c) $H(\theta)$, $Z(\theta)$, and $K$  are the same as in (b)
: $\phi_0 \approx 0.769$, $A=0.5$.
(d) gap-junction coupled Morris-Lecar oscillators 
\cite{rinzel_bookchapt}: $I_{ext,1} = I_0+\Delta I$ and $I_{ext,2} = I_0-\Delta I$ with $\Delta I=0.2$, $A_V = 40$. (top) type I case, $I_0=50$ and $g_{syn} = 0.015$, $\phi_0 \approx 1.279$. (bottom) type II case, $I_0 = 94$ and $g_{syn}=0.01$, $\phi_0 \approx 0.778$.
}
\label{fig_prc_nonoddH}
\end{figure}
%*************************************************

When $H(\theta)$ is not an odd function, the even part of $H$ affects
the dynamics of $\Phi$ and thus the phase shift $\Delta \Phi$ through
Eq. (\ref{eq_phicm}).  
Finding the PRC in the analytic form is not possible for these cases
since we have to solve equation (\ref{eq_phid}) for general initial
data.  Instead, we can get an approximation of the PRC  
in the limit of small changes in $\phi$. 
Let $\phi = \phi_0 + q$ with $|q| \ll 1$. We can linearize Eq. (\ref{eq_phid}) and obtain approximation $q \approx q_0 e^{-2 K {H_{o}}'(\phi_0) (t-t_1)}$ for $t > t_1$ 
where $q_0$ is the change in $\phi$ right after the impulse: $q_0 = A Z_d(\theta_1(t_1), \theta_2(t_1))$. As $t \rightarrow \infty$, $\phi$ returns to $\phi_0$. Thus, $\Delta \phi = 0$.  The phase shift $\Delta \Phi$ is given by 
$\Delta \Phi = A Z_{av}(\theta_1(t_1), \theta_2(t_1))+\int_{t_1}^\infty K\left [H_{e}(\phi)-H_{e}(\phi_0)\right] dt 
\approx A Z_{av}(\theta_1(t_1), \theta_2(t_1))+ \frac{{H_{e}}'(\phi_0)}{2 {H_{o}}'(\phi_0)} q_0$,
where we use $H_{e}(\phi)-H_{e}(\phi_0) \approx {H_{e}}'(\phi_0) q(t)$.

Therefore, the PRC of the oscillator $1$ is 
\begin{eqnarray}
{Z_{c1}}(\theta_1) &\approx&  A Z_{av}(\theta_1, \theta_1-\phi_0) \nonumber \\ 
&+&  \frac{{H_{e}}'(\phi_0)}{2 {H_{o}}'(\phi_0)} A Z_d(\theta_1, \theta_1-\phi_0). 
\label{eq_prc_nonoddH}
\end{eqnarray}

Figure \ref{fig_prc_nonoddH} shows $Z_{c1}(\theta_1)$ with non-odd
coupling functions. In Fig. \ref{fig_prc_nonoddH}(a), we show the PRC
with the simple type of $H$. While $A Z$ and $A Z_{av}$ are similar, the
obtained PRC for the coupled oscillator is significantly different
from them. The curve from Eq. (\ref{eq_prc_nonoddH}) fits well with
simulation results for the entire range of $\theta_1$. Figure
\ref{fig_prc_nonoddH}(b) shows the results with a $H$ function with
higher order terms. We use small $A=0.2$ for this case and the PRC
from the theory fits well with the simulation result.  
In Fig. \ref{fig_prc_nonoddH}(c), we use the same parameters as in (b)
except $A=0.5$. With the larger $A$, the theory mismatches
significantly for a range of phases, 
 but still gives a relatively
similar shape to the simulations. The overall matching is due
to the fact that $q=0$ at some phases satisfying $Z(\theta) =
Z(\theta-\phi_0)$ and around those phases the theory fits well with
simulation results. 
Figure \ref{fig_prc_nonoddH}(d) shows the PRC of gap-junction coupled
Morris-Lecar oscillators  with slightly different injection
currents \cite{rinzel_bookchapt}: $C\dot V_i = -I(V_i,w_i)+I_{ext, i}
+ g_{syn}(V_j - V_i)+A_{\rm V}\delta(t-t_1)$ with $j=2,1$ for $i=1,2$. 
The details are in Ref. \cite{rinzel_bookchapt}.
 The system is simulated using the
4th-order Runge-Kutta method. For type I and type II cases, the theory
gives good fitting with the simulation results with weak stimulus.

Next, we want to understand PRCs for oscillators coupled to many other
oscillators. 
We study the case with globally coupled oscillators: For $i=1,2,..., N,$ 
\begin{eqnarray}
\dot \theta_i &=& \omega_i + \frac{K}{N} \sum_{j=1}^{N} H(\theta_j-\theta_i)+Z(\theta_i)\cdot A\delta(t-t_1), ~~~~ 
\label{eq_model_0}
\end{eqnarray}
where $N$ is the total number of oscillators and others are as defined in the two oscillator system. 

We introduce similar variables as in two coupled oscillators: the average of the phases $\Phi\equiv \frac{1}{N}\sum_{j=1}^{N}\theta_j$  and the phase $\phi_i\equiv  \theta_i - \theta_M$ of oscillator $i$ relative to the phase of oscillator $M$,
where the subscript $M$ denotes the oscillator which has the average frequency $\bar \omega$.
From the definitions of $\Phi$ and $\phi$, we obtain
$\theta_M =  \Phi - \frac{1}{N} \sum_{j=1}^N \phi_j$.  
Because the PRC for other oscillators can be treated similarly and oscillator $M$ follows closely to the collective behavior of the system, we focus on the PRC $Z_c(\theta_M)$ of oscillator $M$.
As in the case of two coupled oscillators, we get 
\begin{eqnarray}
Z_c(\theta_M ) &=&  \Delta\Phi - \langle \Delta \phi \rangle~\textrm{with}~\langle \Delta \phi \rangle \equiv \frac{1}{N} \sum_{j=1}^N \Delta \phi_j.~~~~
\label{eq_kura_Zc}
\end{eqnarray}
The equation for $\Phi$ is
\begin{eqnarray}
\dot \Phi &=& \bar \omega + \frac{K}{N^2} \sum_{i,j=1}^{N} H_{e}(\phi_j-\phi_i) +Z_{av}\cdot A\delta(t-t_1),~~~~~  
\label{eq_kura_Phi}
\end{eqnarray}
where $Z_{av}(\theta_1,\dots, \theta_N) \equiv \frac{1}{N} \sum_{i=1}^N Z(\theta_i(t))$.

For an odd function $H(\theta)$, 
$\Delta \Phi = A Z_{av}$. 
But for a non-odd function $H(\theta)$, the second term contributes to $\Delta \Phi$ and it is not easy to calculate $\Delta \Phi$ analytically.

Let us consider fully locked states first.
For a fully locked state with relative phases $\phi_{i0}$, the
system returns to the locked state after the stimulation and the
relative phase $\phi_{i0}$ can be changed to the equivalent phase
$\phi_{i0} + 2n_i\pi$, where $n_i$ is an integer. Thus, $\Delta \phi_i
= 2 n_i \pi$.  

With $H(\theta) = \sin(\theta+\beta)$, which is a good approximation for many general coupling functions, Eq. (\ref{eq_model_0}) becomes 
\begin{eqnarray}
\dot \theta_i &=& \omega_i + K R \sin(\Theta-\theta_i+\beta)+Z(\theta_i)\cdot A\delta(t-t_1),~~~~~~ 
\label{eq_kura_model}
\end{eqnarray}
where $R$ and $\Theta$ are the order parameter and the corresponding collective phase respectively defined by 
$R e^{i \Theta} \equiv \frac{1}{N} \sum_{j=1}^N e^{i \theta_j}$.
In the frame rotating with the synchronization frequency $\Omega$, the
equation becomes 
\begin{eqnarray}
\dot \psi_i &=& \omega_i - \Omega + K R \sin(\tilde \Theta-\psi_i+\beta) \nonumber \\
&&+Z(\theta_i)\cdot A\delta(t-t_1), 
\label{eq_kura_model2}
\end{eqnarray}
where $\psi_i \equiv \theta_i - (\Omega t + \Theta_0)$, $\tilde \Theta
\equiv \Theta - (\Omega t +\Theta_0)$ and the constant $\Theta_0$ is
chosen such that the stationary value of $\tilde \Theta$ before the
impulse is zero. Let $R_0$ denote the stationary value of $R$. 
We can analyze the stationary state of the system, using self-consistency argument and find $R_0$ and $\Omega$ \cite{kura,sakaguchi1986}. 

For fully locked states or partially locked states where the oscillator $M$ is locked with a locking phase ${\psi_M}^*$ and the oscillators form a stationary distribution relative to the frame rotating with $\Omega$, 
$Z_{av} = \frac{1}{N} \sum_{i=1}^{N} Z(\theta_M+\psi_i-{\psi_M}^*)$, where ${\psi_M}^* = \sin^{-1}(\frac{\bar \omega - \Omega}{K R_0})+\beta$. 
For $Z(\theta) = a_1 + a_2 \sin(\theta +\xi)$, 
using $R_0= \frac{1}{N} \sum_{j=1}^{N}{\rm e}^{{\rm i}\psi_j}$, we obtain 
\begin{eqnarray}
Z_{av} &=& a_1+ a_2 R_0 \sin(\theta_M+\xi-{\psi_M}^*).
\label{eq_kura_Zav}
\end{eqnarray}
Note that since $R_0 < 1$, the magnitude of the sinusoidal part of $Z_{av}$ is smaller than that of $Z$ unless synchrony is perfect.

%*************************************************
%Fig. 3
%\psdraft
\begin{figure}[tb]
\centering
\epsfig{figure=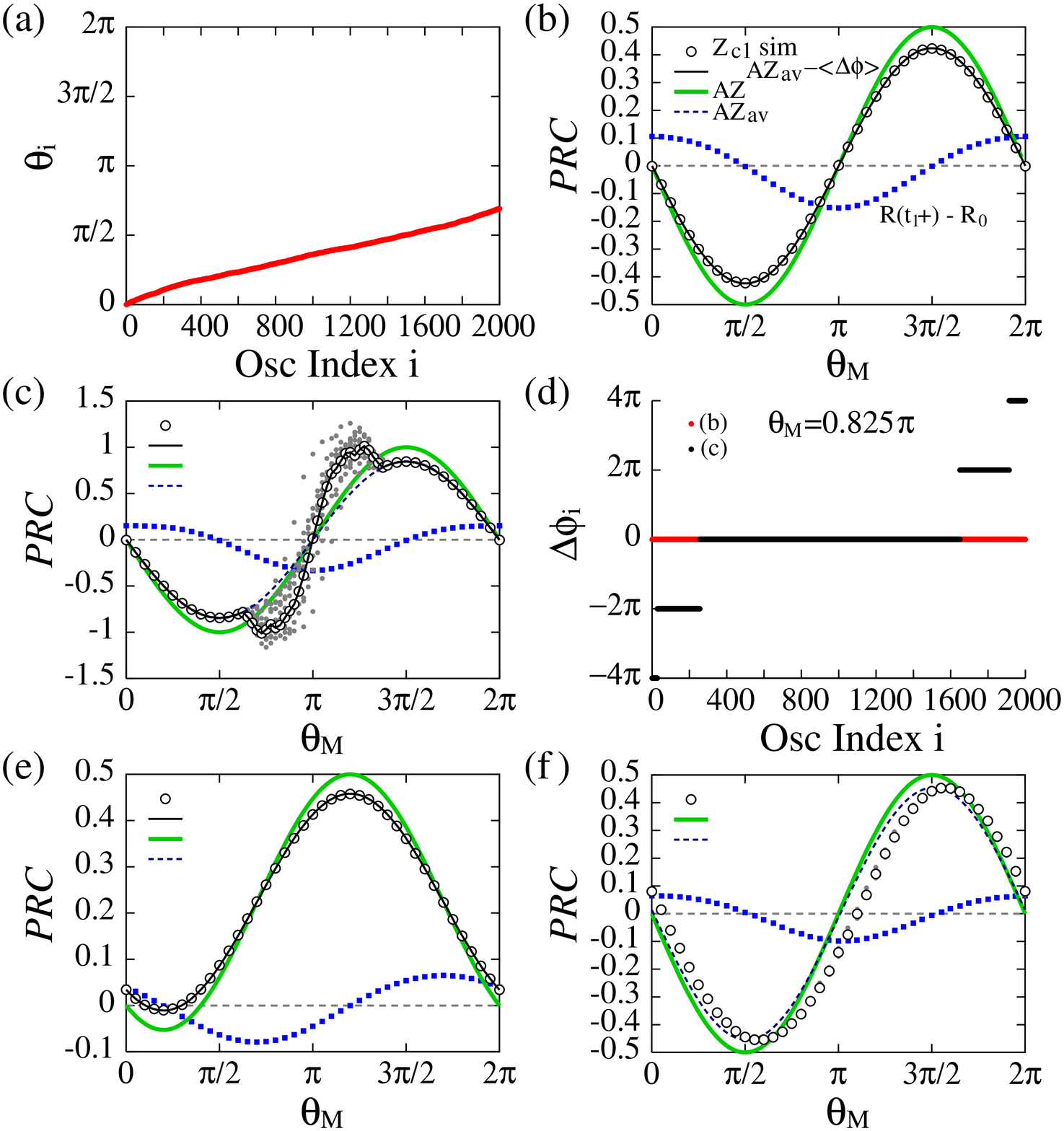, width= 7.5cm}
\caption{(Color online) 
Fully locked cases. 
(a) snapshot of phases of oscillators: a fully locked state  
(b) $A=0.5$. (c) $A=1.0$. (d) $\Delta \phi_i$ for (b) and (c) with $\theta_M=0.825\pi$. 
For (a)-(d), $H(\theta) = \sin\theta$ and $Z(\theta) = -\sin\theta$.
(e) $H(\theta) = \sin\theta$, $Z(\theta)=-\left[\sin(\theta+0.3\pi)-\sin(0.3\pi)\right]/\left[1+\sin(0.3\pi)\right]$, and $A=0.5$.
(f)  $H(\theta) = \sin(\theta+0.1\pi)$, $Z(\theta) = -\sin\theta$, and $A=0.5$.
A uniform distribution is used for $\{\omega_i\}$: (i) $\omega_i = \bar \omega-\Delta \omega+\frac{2\Delta \omega(i-1)}{N-1}$ or (ii) randomly selected $\omega_i$ from $[\bar \omega-\Delta \omega, \bar \omega+\Delta\omega]$. (i) is used for the symbols of (a)-(e) except the gray circles in (b), (c), (e), and (f).  $\bar \omega=\pi/2$.  $\Delta\omega=0.6$.  $K=0.8$ for (a)-(e) and $K=1$ for (f). 
}
\label{fig_prc_kura_ud}
\end{figure}
%*************************************************

Figures \ref{fig_prc_kura_ud}(a)-(e) show results with  $H(\theta)=\sin\theta$ and a uniform distribution for the frequencies of the oscillators.
We use  $Z(\theta)=-\sin\theta$ for (a)-(d), and an asymmetric $Z(\theta)$ for (e).
With the given coupling strength, the system shows a fully locked
state (Fig. \ref{fig_prc_kura_ud}(a)). Figures
\ref{fig_prc_kura_ud}(b) and (c) show the PRCs for different values of
$A$. With weak stimulation (Fig.  \ref{fig_prc_kura_ud}(b)), all $\phi_i$ return to the unperturbed values ($\Delta\phi_i=0$ for all $i$, Fig. \ref{fig_prc_kura_ud}(d)) and the PRC is shown to be contributed only by $\Delta \Phi = AZ_{av}$. 
The prediction $Z_c(\theta_M)=AZ_{av}$ from the theory (Eq. (\ref{eq_kura_Zav})) fits well with the simulation results.  In contrast, 
with a stronger impulse (Fig. \ref{fig_prc_kura_ud}(c)), the simulation results deviate from $Z_c(\theta_M)=AZ_{av}$ for some range of $\theta_M$. We calculate $\langle \Delta \phi \rangle$ from the simulations and it accounts for the deviation as predicted from Eq. (\ref{eq_kura_Zc}). 

The deviation in Fig. \ref{fig_prc_kura_ud}(c) can be understood as follows.
Nonzero $\Delta \phi_i$ can occur only when the order parameter transiently decreases.  For $\theta_M \in (\pi/2,
\pi)$, the impulse disperses the locked group, because the trailing
oscillators receive more negative impact than the leading ones.
 Thus, the order parameter decreases from $R_0$ to $R({t_1}+)$ ($< R_0$). 
$R$ can decrease more depending on the behavior
of the oscillators, and then returns to $R_0$.  
The collective phase $\Theta$ also decreases ($\tilde \Theta({t_1}+) < 0$), because most of the phases of the oscillators decrease due to the impulse.
The sudden changes in $R$ and $\tilde \Theta$ affect the dynamics of
oscillators. The behaviors of oscillators right after the impulse can be
described by the equation $\dot \psi_i = \omega_i - \Omega + K
R({t_1}+) \sin(\tilde \Theta({t_1}+) -\psi_i)$ with $\psi_i({t_1}+) =
{\psi_i}^* + Z(\theta_M(t_1) + {\psi_{i}}^*-{\psi_M}^*)$, where ${\psi_i}^*$ is the locking phase for oscillator $i$. 
The trajectory of oscillator $i$ can escape completely from the basin of attraction of $\phi_{i0}$ during the transient behavior of $R$ and settle to the equivalent phases $\phi_i=\phi_{i0}+2n_i\pi$.
Since the curves for $(\psi_i, \dot \psi_i)$ are shifted to the left
due to the negative $\tilde \Theta({t_1}+)$ and upwards(downwards) for
the oscillators with $\omega_i > \Omega$ ($\omega_i < \Omega$), the
oscillators with frequencies far from the average one can escape and
those with higher frequencies escape first. Because of
this, $\langle \Delta \phi \rangle > 0$ 
and the PRC deviates
negatively from $A Z_{av}$ (Eq. (\ref{eq_kura_Zav})).  
The oscillators with frequencies far from the average one have more
chance to have higher $n$ (Fig. \ref{fig_prc_kura_ud}(d)), because they can drift faster and stay
unlocked longer. Other ranges of $\theta_M$ can be understood similarly. 
%*************************************************
%Fig. 4
%\psdraft
\begin{figure}[tb]
\centering
\epsfig{figure=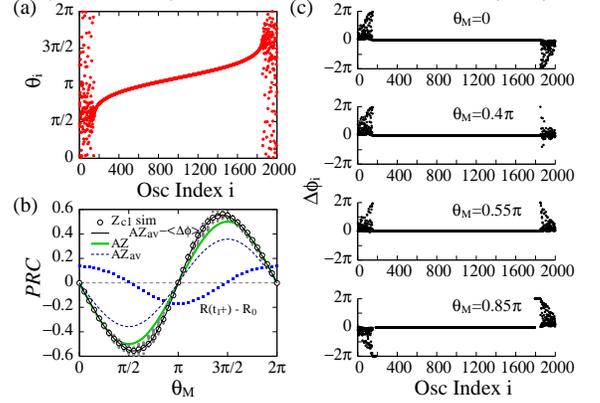, width= 7.5cm}
\caption{(Color online) 
Partially locked cases with $H(\theta) = \sin\theta$. 
(a) snapshot of phases of oscillators: a partially locked state.  
(b) $A=0.5$. (c) $\Delta \phi_i$.  
$Z(\theta) = -\sin\theta$.
$\{\omega_i\}$ obeys a Gaussian distribution $g(\omega, \bar \omega)=\frac{1}{\sigma \sqrt{2\pi}}\exp({-\frac{(\omega-\bar\omega)^2}{{2\sigma^2}}})$: (i) 
$\omega_{(N/2)+k} = \bar \omega + y_k$, $\omega_{(N/2)-k+1} = \bar
\omega - y_k$ with $y_k = (x_{k-1} + x_k)/2$, $x_{k+1}=x_k +
N^{-1}/g(x_k,0)$, and $x_0 = 0$ for $k=1,..., N/2$ \cite{daido2000} or
(ii) $\omega_i = \bar \omega - y_i$ and $\omega_{i+N/2} = \bar \omega
+ y_i$ with $i \leq N/2$ and $y_i$ randomly selected according to
$g(y,0)$ with $y>0$. (i) is used for the simulations(symbols) of
(a)-(c) except the gray circles in (b).  $\bar \omega=\pi/2$,
$\sigma=0.3$, and  $K=0.6$.  
%-------
}
\label{fig_prc_kura_gd}
\end{figure}
%*************************************************

When $H(\theta)$ is not odd, it is difficult to find the general
results. For $H(\theta) = \sin(\theta+\beta)$, we can see that the
second term of Eq. (\ref{eq_kura_Phi}) is equal to $K R^2 \sin\beta$.   
Thus, $\Delta \Phi = A Z_{av} + K\sin\beta \int_{t_1}^{\infty} (R^2-{R_0}^2) dt$. 
With weak stimulus, $Z_c(\theta_M) = \Delta \Phi$ and the PRC deviates positively (negatively) from $A Z_{av}$ for values of
$R({t_1}+) > R_0$ ($R({t_1}+) < R_0$) (Fig. \ref{fig_prc_kura_ud}(f)).  
The values of $R(t_1+)$ are easily calculable using $Z$ and the distribution for $\omega$. 

Finally, let us briefly consider partially locked cases in the limit
of $N \rightarrow \infty$.  
When the system exhibits a partially locked state, the drifting
oscillators form a stationary distribution in the frame rotating with
the synchronization $\Omega$. In the original frame, we can say that
the distribution rotates with $\Omega$.  We can define PRCs for locked
oscillators. 
Let us consider cases with $H(\theta) = \sin\theta$. Figure
\ref{fig_prc_kura_gd}(b) shows the PRC $Z_c(\theta_M)$ of oscillator
$M$ for the partially locked state of (a).  
We can understand $Z_c(\theta_M)$ through Eq. (\ref{eq_kura_Zc}).
Since $H$ is an odd function, $\Delta \Phi = A Z_{av}$ (Eq. (\ref{eq_kura_Zav})). 
While for the locked
oscillators $\Delta \phi_i = 2n_i \pi$ and is nonzero in some ranges
as in the fully locked cases, for the drifting oscillators $\Delta
\phi_i$ is usually not an integer multiple of $2\pi$ and can be
nonzero in any ranges (Fig. \ref{fig_prc_kura_gd}(c)). Simulations
show that $\Delta \phi_i$ for the drifting oscillators contribute
significantly to the PRC and the PRC (symbols,
Fig. \ref{fig_prc_kura_gd}(b)) differ from $A Z_{av}$ (the dashed
curve) for almost the entire range of $\theta_M$. 

In summary, we have investigated the PRCs of coupled oscillators in terms of the PRCs of individuals, the nature of the coupling, and the relative phases of the oscillators.    
Our approach of obtaining PRCs using the average and relative phases can be applicable to oscillators on different type of networks.

This work was supported by National Science Foundation grant DMS05135.

%==================================================================


\begin{thebibliography}{99}
\bibitem{pikovsky_book} A. Pikovsky, M. Rosenblum, and J. Kurths, {\it Synchronization: a universal concept in nonlinear sciences} (Cambridge University Press, Cambridge, 2001)

\bibitem{kura} Y. Kuramoto, {\it Chemical Oscillations, Waves, and Turbulence} (Springer, Berlin, 1984); S. H. Strogatz, Physica D {\bf 143}, 1 (2000); J. A. Acebr{\'o}n {\it et al.}, Rev. Mod. Phys. {\bf 77}, 137 (2005).

\bibitem{winfree} A. T. Winfree, {\it The Geometry of Biological Time}, 2nd ed.  (Springer, New York, 2001).

\bibitem{rinzel_bookchapt} J. Rinzel and G. B. Ermentrout, in {\it Methods in Neuronal Modeling}, 2nd ed. (MIT Press, Cambridge, MA, 1998).
% {\it Analysis of Neural Excitability and Oscillation}

\bibitem{ermentrout2001} G. B. Ermentrout and D. Kleinfeld, Neuron {\bf 29}, 33 (2001).

\bibitem{tass_book} P. Tass, {\it Phase Resetting in Medicine and Biology} (Springer, Berlin, 1999).

\bibitem{prc_papers} 
G. B. Ermentrout, Neural Comp. {\bf 8}, 979 (1996);
L. Glass, Y. Nagai, K. Hall, M. Talajic, and S. Nattel, Phys. Rev. E {\bf 65}, 021908 (2002); E. Brown, J.  Moehlis, and P. Holmes, Neural Comp. {\bf 16}, 673 (2004); R. F. Gal{\' a}n, G. B. Ermentrout, and N. N. Urban, Phys. Rev. Lett. {\bf 94}, 158101 (2005); R. Gunawan and F. J. Doyle III, Biophys. J. {\bf 91}, 2131 (2006); E. M. Izhikevich, {\it Dynamical Systems in Neuroscience} (MIT Press, Cambridge, MA, 2007).  

\bibitem{kawamura2008} Y. Kawamura {\it et al.}, Phys. Rev. Lett. {\bf 101}, 024101 (2008). 


\bibitem{sakaguchi1986} H. Sakaguchi and Y. Kuramoto, Prog. Theor. Phys. {\bf 76}, 576 (1986).
% A soluable active rotator model showing phase transitions via mutual entrainment

\bibitem{daido2000} H. Daido, Phys. Rev. E {\bf 61}, 2145 (2000).
\end{thebibliography}
\end{document}